\documentclass[11pt,twoside]{article}

\usepackage{asp2006}
\usepackage{epsf}
\usepackage{psfig}
\usepackage{graphicx}
\usepackage{lscape}
\usepackage{amsbsy}
\usepackage{amssymb}

\begin{document}
\title{Why are some A stars magnetic, while most are not?}

\author{G.A. Wade, J. Silvester, K. Bale, N. Johnson, J. Power}   
\affil{Royal Military College of Canada}    
\author{M. Auri\`ere, F. Ligni\'eres, B. Dintrans, J.-F. Donati, A. Hui Bon Hoa, D. Mouillet, S. Naseri, F. Paletou, P. Petit, F. Rincon, N. Toque}
\affil{Observatoire Midi-Pyr\'en\'ees}    
\author{S. Bagnulo, C.P. Folsom}
\affil{Armagh Observatory}
\author{J.D. Landstreet}
\affil{University of Western Ontario}
\author{M. Gruberbauer, T. Lueftinger}
\affil{Insit\" ut fur Astronomie, Wien}
\author{S. Jeffers}
\affil{Universiteit Utrecht}
\author{A. L\`ebre}
\affil{GRAAL, UniversitŽ Montpellier II}
\author{S. Marsden}
\affil{Anglo-Australian Observatory}

\begin{abstract} 
  
A small fraction of intermediate-mass main sequence (A and B type) stars have strong, organised magnetic fields. The large majority of such stars, however, show no evidence for magnetic fields, even when observed with very high precision. In this paper we describe a simple model, motivated by qualitatively new observational results, that provides a natural physical explanation for the small fraction of observed magnetic stars.
  
\end{abstract}


\section{Introduction}   

In the sun and essentially all other low-mass stars, vigorous magnetic activity results from the cyclical conversion of convective and rotational mechanical energy into magnetic energy, generating highly structured and variable magnetic fields whose properties correlate strongly with stellar mass, age and rotation rate. Although the dynamo mechanism which drives this process is not understood in detail, its basic principles are well established.

On the other hand, some higher-mass stars (mainly the so-called peculiar A-type stars, or Ap stars) also host magnetic fields. These fields differ from those of low-mass stars in a number of important ways: they are detected in only a small fraction of stars, they are topologically much simpler and often much stronger, their characteristics show no clear correlation with other stellar properties, and they remain essentially unchanged over relatively long astrophysical timescales. 

The striking differences between the characteristics of the magnetic fields of low-mass and those of higher-mass stars argues strongly for fundamentally different physical origins. 

Although a variety of models have been proposed to explain the magnetic fields of intermediate-mass stars, the weight of opinion currently supports a {\em fossil} origin: that the observed magnetic fields are the slowly-decaying remnants of field swept up from the interstellar medium during the process of star formation, or possibly generated by a pre-main sequence dynamo which has since ceased to operate. 

Such a model has the advantages that the resultant stellar magnetic fields need not show any obvious correlation with stellar properties, and that the fields should have simple, stable topologies (as the timescale for Ohmic decay of the field decreases as the square of the characteristic field length scale, and the decay timescale for the dipole component is approximately $10^9-10^{10}$~years). However, a significant weakness of the fossil model has been its inability to explain why only a few percent of higher-mass stars show fields, whereas the large majority do not (even when observed with very high observational precision; see e.g. Shorlin et al. 2002).

In this paper, we review recent observations (Auri\`ere et al. 2007) showing the existence of a field strength lower threshold in the Ap stars. Based on this qualitatively new observational result, we propose a simple model that naturally explains this "magnetic dichotomy" within the context of the fossil model.

\section{Weak magnetic fields in Ap stars}

Auri\`ere et al. (2007) undertook a systematic monitoring of the magnetic fields of 28 bright Ap stars presumed to have very weak magnetic fields. Although magnetic fields had never generally been detected in the targets selected by Auri\`ere et al. (despite previous attempts), those stars were predicted to be magnetic based on the distinctive spectral peculiarities and variability that identified them as Ap stars. Ultimately, they succeeded in detecting the longitudinal Zeeman effect in every one of their targets, leading to their first basic conclusion: that, as had previously been assumed, all confidently spectroscopically-classified Ap stars, when observed with sufficient precision and tenacity, show evidence for organised magnetic fields. Examples of Stokes $V$ signatures observed by Auri\`ere et al. (2007) in line profiles of previously-undetected Ap stars are shown in Fig. 1.

\begin{figure*}[!ht]
\plottwo{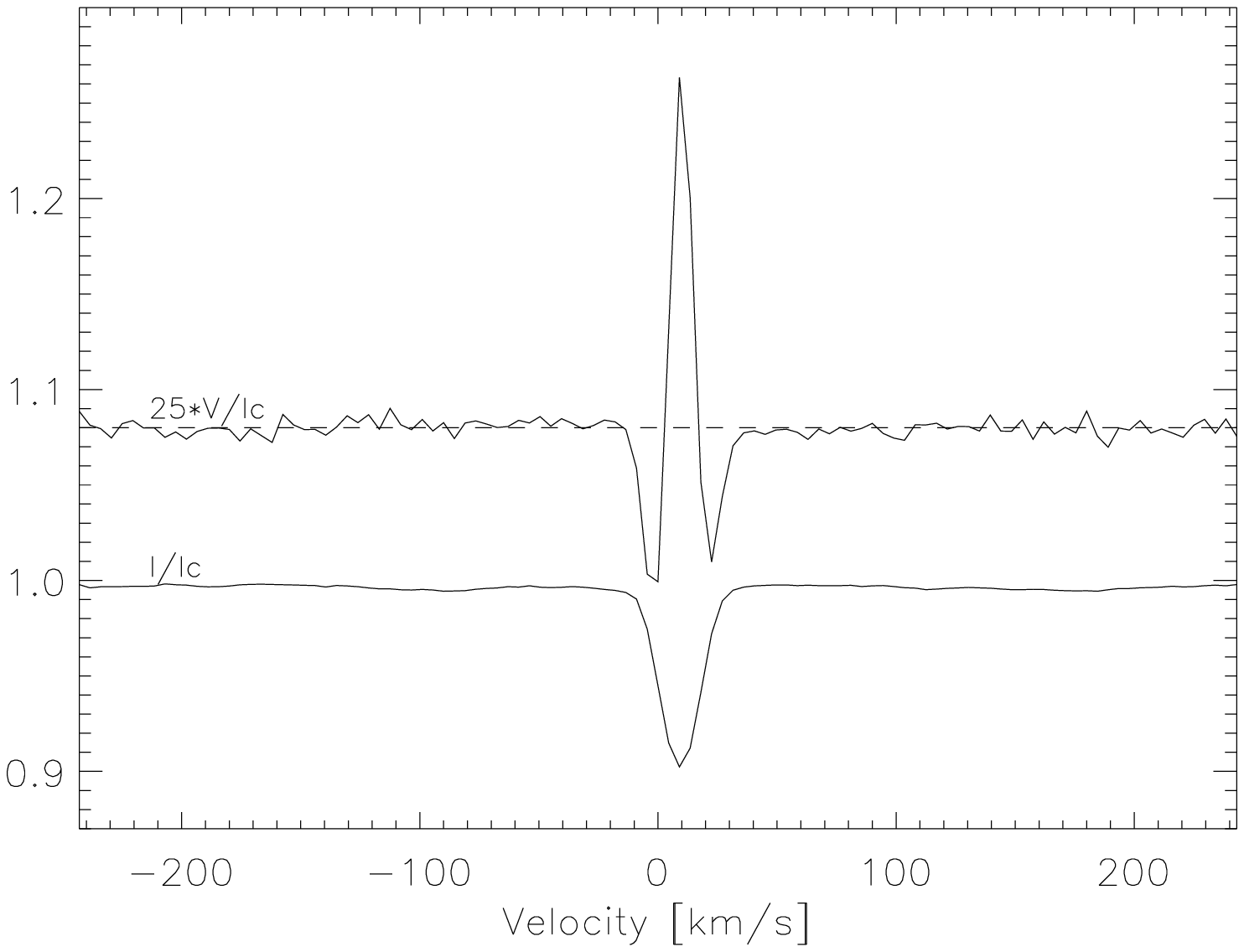}{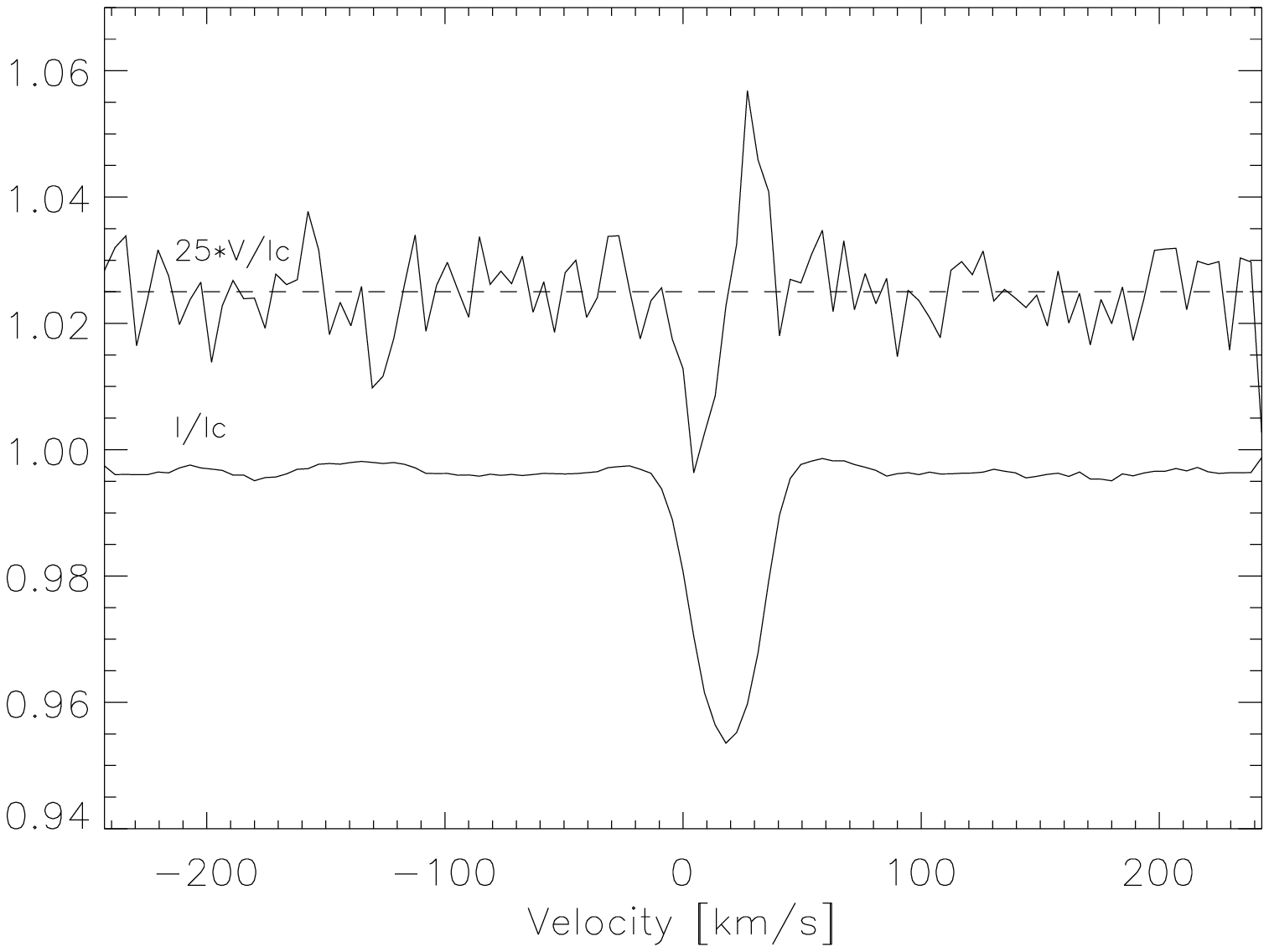}
\caption{Stokes $I$ and $V$ Least-Squares Deconvolved line profiles of the Ap stars HD 43819 (left) and HD 171782 (right). Stokes $V$ Zeeman signatures (magnified and shifted for display purposes) are clearly detected. Adapted from Auri\`ere et al. (2007).}
\label{aa_mrf}
\end{figure*}

\section{The magnetic dichotomy}

Demonstrating the existence of a magnetic dichotomy relies not only on establishing the presence of fields in the Ap stars, but also on showing confidently that no fields are present in the non-Ap stars. Although a variety of searches for magnetic fields in non-Ap intermediate mass stars have been conducted, the most recent and comprehensive are those by Shorlin et al. (2002) and Bagnulo et al. (2006). Shorlin et al. used the high-resolution spectropolarimeter MuSiCoS to search for Stokes $V$ Zeeman signatures in spectra of 63 non-Ap intermediate-mass stars, finding no evidence of magnetic fields, with a median longitudinal field formal error of just 22~G. (For comparison, the median error achieved by Auri\`ere et al. was 40~G.) Bagnulo et al. used the low-resolution FORS1 spectropolarimeter to measure magnetic fields of a large sample of intermediate-mass stars in open clusters. In their sample of 138 non-Ap stars, no magnetic field was detected, with a median longitudinal field error bar of 136~G. These results are interpreted as indicating that organised magnetic fields, similar to those observed in the Ap stars, are not present in non-Ap intermediate-mass stars. 

\section{Discovery of a field strength lower threshold}

Using their 282 new measurements of the longitudinal magnetic field, Auri\`ere et al. (2007) determined the magnetic dipole characteristics (the dipole polar strength $B_{\rm d}$ and obliquity angle $\beta$) of 24 of their target Ap stars, in the context of the Oblique Rotator Model. (Although models are missing for 4 stars due to insufficient data to characterise the rotational variation of the longitudinal field, these stars were clearly detected in their survey). The distribution of inferred dipole strengths is shown in Fig. 2 (left panel). 

The most remarkable characteristic of this distribution is the near-complete absence of stars with dipole strengths below about 300 G. This result cannot be due to a detection threshold effect, because the magnetic field was detected for every star in their sample. Nor is it likely that it is due to a selection effect (e.g. the systematic exclusion of stars with $B_{\rm d}<300$~G) , as their sample was constructed specifically to include the Ap/Bp stars with the weakest magnetic fields.

What is clearly demonstrated by the results of Auri\`ere et al. (2007)  is that the number of Ap/Bp stars does not continue to increase monotonically to arbitrarily small field strengths. Instead, it appears to plateau around 1~kG, and to decline rapidly below about 300~G.  This conclusion is supported by the results of Power et al. (in preparation), who employed similar techniques to characterise the dipole field strengths of Ap stars within 100 pc of the sun. Their histogram of dipole fields strengths for 31 stars (Fig. 2, right panel) shows an similar absence of stars with $B_{\rm d}$\ ${\mathrel{\hbox{\rlap{\hbox{\lower4pt\hbox{$\sim$}}}\hbox{$<$}}}}$\ 300~G.

Auri\`ere et al. also point out that the surface field intensities $B_{\rm d}$ they derive are probably themselves only lower limits on the true surface field strength. When sufficient data are available, detailed models of magnetic Ap stars nearly always show evidence of higher-order multipolar contributions to the magnetic field (e.g. Landstreet 1988, Landstreet \& Mathys 2000). These higher-order field components contribute only weakly to the longitudinal field variation, although they can have surface intensities comparable to that of the dipole component. 

A straightforward interpretation of the behaviour observed in Fig. 2 is that there exists a minimum magnetic field strength necessary for the generation of the characteristic chemical peculiarities and variability of Ap stars. 

\begin{figure*}[!ht]
\plotfiddle{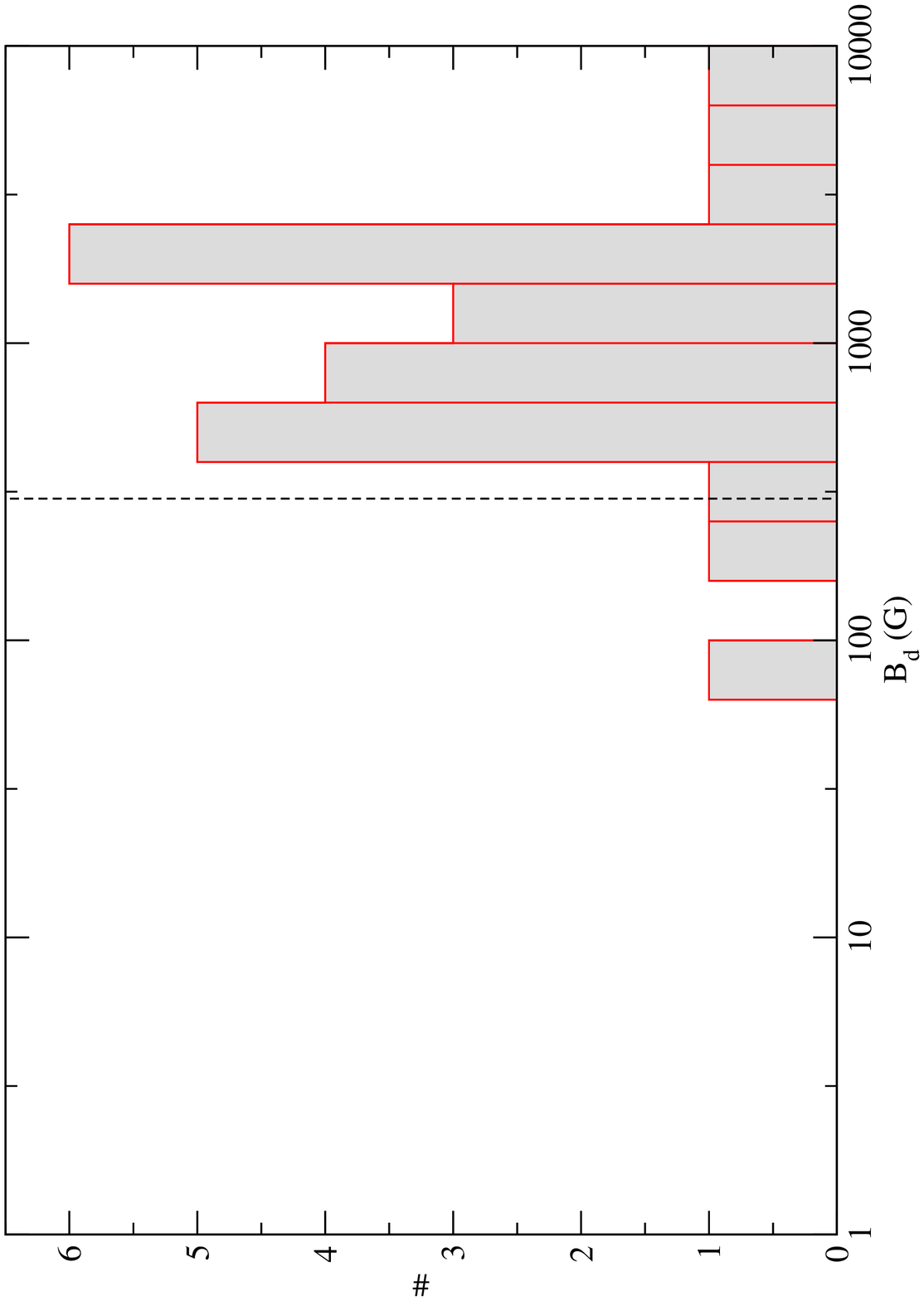}{4cm}{-90}{25}{25}{-197}{+130}
\plotfiddle{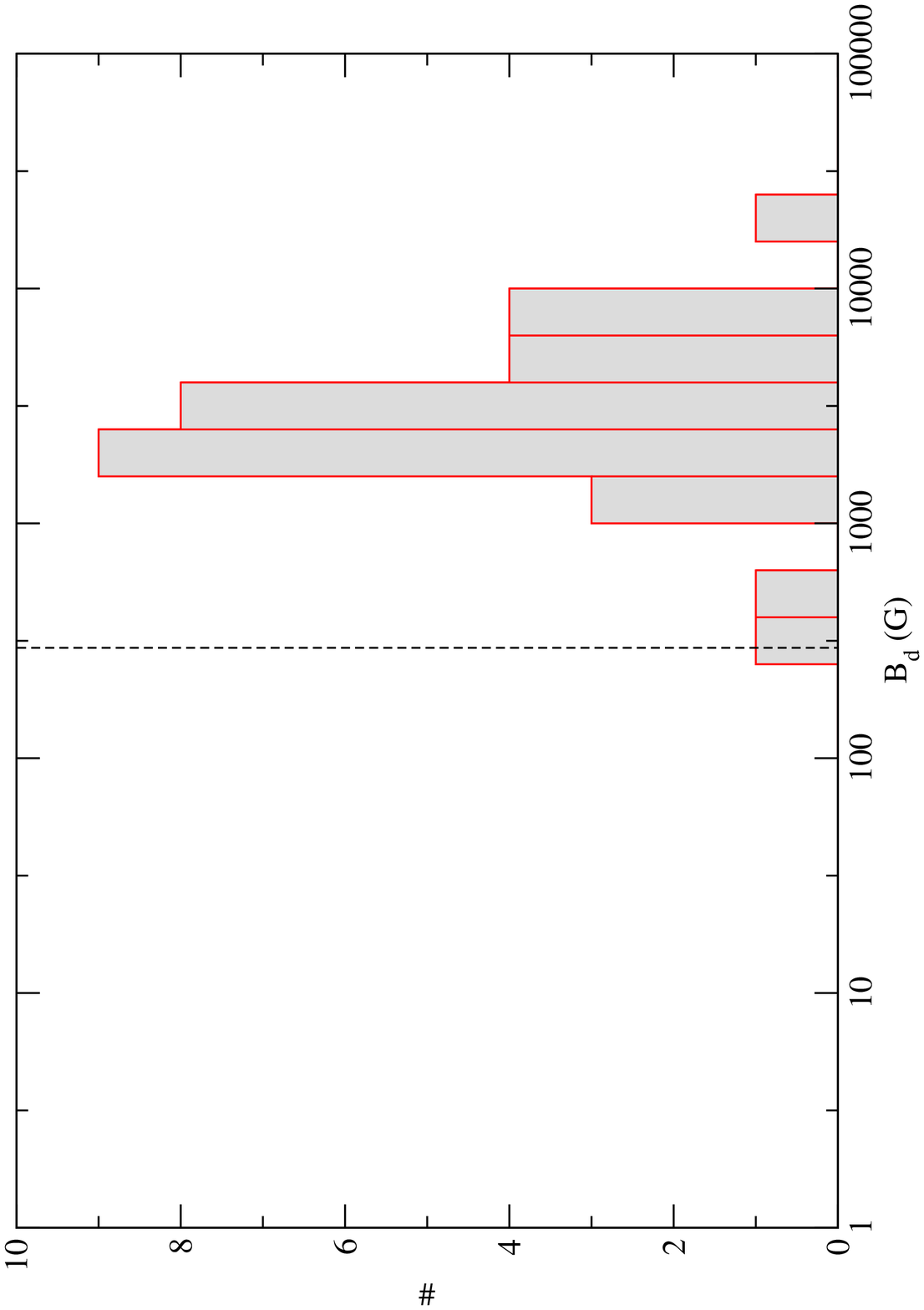}{0cm}{-90}{25}{25}{-13}{+154}
\caption{Histograms of derived surface dipole magnetic fields strengths for stars observed by Auri\`ere et al. (2007 - bright Ap stars with presumably weak magnetic fields; left panel) and Power et al. (in prep. - Ap stars within 100 pc of the sun; right panel).  The dashed lines correspond to $B_{\rm d}=300$~G. Note the difference in the horizontal scales.}
\label{aa_mrf}
\end{figure*}

\section{The lower threshold as a critical field strength}

The 100{\%} Zeeman detection rate obtained by Auri\`ere et al. (2007) strongly suggests that all Ap/Bp stars host detectable magnetic fields. Moreover, it appears that a threshold magnetic field of about 300 G exists, below which fields are very rare, and perhaps altogether absent.

An interpretation of this result is that there exists a critical field strength (corresponding to the observed field threshold) above which stable magnetic configurations can exist
and below which any large scale field configuration is destroyed due to an instability.
The instability is expected to generate opposite polarities at small length scales, thus
strongly reducing the magnitude of the integrated longitudinal field through cancellation effects, and accelerating Ohmic decay.
For a sample of stars containing both stable and unstable field configurations,
this scenario would imply a strong jump in the measured values of the longitudinal fields
or a lower bound of the magnetic field, depending on the detection limit.

The existence of stable large scale magnetic fields in stars is primarily
supported by observations of the magnetic fields of Ap stars and white dwarfs.
Theoretically, although no stable field configuration is known in an analytical form, it has been proposed that the
combination of azimuthal and poloidal field might be stable, as recent numerical simulation tend to confirm (Braithwaite \& Spruit, 2004).
However, when the magnetic field is sufficiently weak to be wound up by differential rotation, the resulting field, predominantly azimuthal with respect to the rotation axis, can be subject to various instabilities. As recently reviewed by Spruit (1999), the most vigorous
of these instabilities is a pinch-type instability
first considered in a stellar context
by Tayler (1973).
Auri\`ere et al. (2007) estimate the critical magnetic field below which the winding-up process induces an instability
and above which the action of magnetic torques on the differential rotation limits the winding-up before the instability sets in.
The winding-up time scale is $t_w = 1/ (q \Omega )$ where
$q = r \| \nabla \Omega \| / \Omega = r / \ell $ is a dimensionless measure of the differential rotation. The winding-up of the axisymmetric part of the original poloidal field $\vec{B}_{\rm p}^{\rm sym}$ by the differential rotation being governed by
${\partial}_t B_{\phi} = r \sin \theta \vec{B}_{\rm p}^{\rm sym} \cdot \vec{\nabla} \Omega $, the time scale $t_w$ corresponds more specifically to the time necessary
to produce an azimuthal field component $B_{\phi}$ of the same amplitude as $\vec{B}_{\rm p}^{\rm sym}$.
On the other hand, Lorentz forces will affect the differential rotation after a Alfv\'en travel time
calculated on the shear length scale $\ell$, that is  $t_A = \ell  (4 \pi \varrho )^{1/2} / B$.
Equating both time scales gives a local  order of magnitude estimate of the critical magnetic field, $B_c \simeq (4 \pi \varrho)^{1/2} r \Omega $. Its value
can be expressed in terms of the equipartition field of a typical Ap star as follows:

\begin{equation}
\frac{B_c}{B_{eq}} \simeq 2 \left( \frac{P_{rot}}{5_{day}} \right)^{-1} \left( \frac{r}{3 R_\odot} \right) \left( \frac{T}{10^4 K} \right)^{-1/2}.
\end{equation}

As $B_{eq} \simeq 170$ G at the surface ($\tau_{5000} = 2/3$) of a typical Ap star  ($\log g=4, T_{\rm eff} = 10^4$~K)
 the derived critical field is close to the observed 300 G threshold. Calculation of the critical field $B_c$ for each star in the samples of Auri\`ere et al. and Power et al. also 
shows that all stars satisfy $B_d > B_c$ within the uncertainties. Moreover, the magnetic fields of all stars 
with short rotational period (under 2~d) are compatible
with the dependence of $B_c$ on $P_{rot}$ $(B_c \propto P_{{rot}}^{-1})$, as their 
dipolar fields are substantially
greater than the sample median field strength.
It is however important to stress that although the local order of 
magnitude estimate of $B_c$ is consistent with the present
observational data, a detailed and non-local modeling is required to 
specify the critical field strength below which
differential rotation destabilizes large scale field configurations.
Note that the threshold value of the magnetic field is also higher than the magnetic field threshold necessary to trigger the Tayler instability, according to the criterion given
by Spruit (1999).
The large scale field is then destroyed by the development of the instability.
An example of the non-linear evolution of such unstable configurations has been recently considered in a solar context (Brun \& Zahn 2006) confirming
that the resulting field is structured on small latitudinal length scales.

The above scenario can thus qualitatively explain the existence of an apparent lower bound in the strength of magnetic fields of Ap stars. 
By extension, such a model could provide a basis to explain why magnetic fields are observed in only a small fraction of intermediate-mass stars. If the initial magnetic field strength probability distribution of intermediate-mass stars increases (say) exponentially toward weak fields, the large majority of A-type stars, after formation, would have fields weaker than the critical field described by Eq. (1). The fields of such stars would be unstable and decay; they would therefore appear at the main sequence showing no evidence of a magnetic field.   

Another advantage of the scenario described above is that it may also naturally explain the even greater rarity of magnetic fields detected in massive stars. For a typical main sequence A0p star (with $P=5$~d, $R=3~R_\odot$ and $T_{\rm eff}=10000$~K), Eq. (1) yields $B_c\simeq 2\, B_{eq}\sim 300$~G. However, for a main sequence B0p star (with $T_{\rm eff}=31000$~K, $R=7.2~R_\odot$ and $P=2$~d), $B_c\simeq 7 B_{eq}\sim 2$~kG. With a substantially larger critical field strength, massive stars are substantially less likely to retain their magnetic fields (assuming an initial field probability distribution similar to that of Ap stars, decreasing toward strong fields).

\acknowledgements 
GAW and JDL are supported in part by a Discovery Grant from the Natural Science and Engineering Research Council of Canada (NSERC).


\end{document}